# Brillouin-zone spectroscopy of nonlinear photonic lattices

Guy Bartal,[1] Oren Cohen,[1] Hrvoje Buljan,[1,2] Jason W. Fleischer,[1,3] Ofer Manela,[1]

Mordechai Segev[1]

[1]Physics Department, Technion - Israel Institute of Technology, Haifa 32000, Israel
[2]Department of Physics, University of Zagreb, PP 322, 10000 Zagreb, Croatia
[3]Electrical Engineering Department, Princeton University, New Jersey 08544

**Abstract**

We present a novel, real-time, experimental technique for linear and nonlinear Brillouin zone spectroscopy of photonic lattices. The method relies on excitation with random-phase waves and far-field visualization of the spatial spectrum of the light exiting the lattice. Our technique facilitates mapping the borders of the extended Brillouin zones and the areas of normal and anomalous dispersion within each zone. For photonic lattices with defects (e.g., photonic crystal fibers), our technique enables far-field visualization of the defect mode overlaid on the extended Brillouin zone structure of the lattice. The technique is general and can be used for photonic crystal fibers as well as for periodic structures in areas beyond optics.

It is very often that the characteristics of waves propagating in periodic structures are best described in Fourier space. While waves in periodic structures occur throughout science, photonic systems have the advantage that optical inputs can be easily manipulated and results can be directly imaged. Moreover, optics provides many technologically important systems relying on the dynamics in periodic structures, such as photonic crystals [1,2,3], photonic crystal fibers (PCF) [4,5,6] and waveguide arrays. The periodicity partitions the linear transmission spectrum of waves propagating in such structures into bands of propagating (Bloch) modes separated by forbidden gaps. The lattice geometry, and the corresponding wave dynamics, is most efficiently presented in the Fourier (momentum) space through the extended Brillouin zone map [7], whose boundaries are defined by the Bragg-reflection planes. For example, Fig. 1a shows the first and second Brillouin zones of a square lattice with the high-symmetry points (Γ, X, and M) marked by white dots. Within each band, the curvature of the transmission spectrum varies with the spatial frequency (transverse momentum) and changes sign. This results in regions of normal and anomalous dispersion/diffraction in exactly the same way as band curvature causes positive and negative effective mass in solid-state physics. The first two bands in the transmission spectrum are shown in Fig. 1b, with the dispersion curves between the symmetry points of the first two bands plotted in Fig. 1c.

The equivalence between optical waves in a periodic dielectric structure and electrons in a periodic atomic potential has opened up the areas of "Bloch wave optics" [8] and "photonic bandgap engineering" [2]. For example, photonic crystals can exhibit a gap in the electromagnetic spectrum: a range of frequencies at which light cannot propagate. In a similar vein, Bragg diffractions from such structures give rise to omni-directional mirrors reflecting all incident waves within a wide angular range, manifesting a gap in the spatial transmission spectrum [9]. These ideas, combined with a dielectric defect embedded in the periodic structures, have allowed the fabrication of photonic crystal fibers (PCFs) [10] and photonic bandgap lasers [11]. In nonlinear photonic lattices, the defect can be self-induced, leading to the formation of lattice solitons [12, 13, 14]. Nonlinearity also serves to couple linear modes, making nonlinear photonic systems excellent platforms to study (image) fundamental problems in lattice dynamics.

Here, we present a novel experimental technique for linear and nonlinear Brillouin zone spectroscopy of photonic lattices. Our technique facilitates a direct visualization of the lattice structure in Fourier space by mapping the borders of the extended Brillouin zones, and marking the areas of normal and anomalous dispersion within them. The method relies on excitation of the lattice modes with partially-incoherent waves, performing an optical Fourier transform and measuring the power spectrum. The technique is general and can be used for any periodic structure. We experimentally demonstrate it on square and hexagonal waveguide arrays, and also study 2D waveguide arrays with positive and negative defects embedded in their structure (equivalent to solid-core and hollow-core photonic crystal fibers, respectively). The method provides a powerful diagnostic tool for photonic lattices. Equally important, our technique points the way towards future experiments on nonlinear mode coupling which have been intriguing scientists since the pioneering work of Fermi, Pasta, and Ulam in 1955 [15].

To map out the momentum space of a lattice, the lattice should be probed by a broad spectrum of its eigen-modes (Bloch waves) in the mapped region of k-space. Preferably, one would like to excite the lattice with a broad and uniform spectrum of Bloch modes, residing in several different bands. At the same time, the probe beam must also be broad enough in real space to sample a large enough number of unit cells. These requirements can be reconciled simultaneously by a random-phase (spatially incoherent) probe beam, which facilitates homogeneous excitation of several BZs with a beam that occupies numerous lattice sites. Moreover, the time-averaged intensity of such an incoherent probe beam, which excites many Bloch modes, is smooth, because the speckles are washed out. For these reasons, we probe the photonic lattices with a partially-spatially-incoherent (random-phase) beam, which has a uniform spatial power spectrum extending over several Brillouin zones, and is broad enough to cover numerous lattice sites.

Our experiments are carried out in optically-induced photonic lattices [13], formed in a highly anisotropic photosensitive nonlinear crystal, utilizing the methods that have recently proven to be very successful for spatial solitons experiments in photonic lattices [13,16]. The experimental setup is shown in Fig. 1d. In all experiments, we use a 488 nm laser beam. The 2D square lattice is induced by interfering two pairs of plane waves in the nonlinear material, while the hexagonal (trigonal) lattice is induced by interfering three plane waves under similar conditions [Fig 2]. In both cases, the lattice-forming waves are polarized so as to propagate linearly in the nonlinear medium, serving only to induce the periodic photonic structure. Our partially-spatially-incoherent probe beam is generated by passing a laser beam through a rotating diffuser, and the power spectrum of the emerging beam is controlled using a filter in the focal plane of a lens. The laser beam exiting the diffuser is imaged onto the input face of the nonlinear crystal in which the photonic lattice is induced. The data is taken by monitoring (photographing) the intensity distribution at the focal plane of the "exit lens" positioned so as to visualize the Fourier power spectrum of the light exiting the crystal.

Consider first the linear scheme for mapping the extended BZ of square and hexagonal lattices, both with ~10 $\mu m$ period. The probe beam is 25 $\mu m$ FWHM, having a transverse correlation distance of 5 $\mu m$. Typical results are shown in Fig. 2, where the upper (lower) row depicts experimental data obtained with a square (hexagonal) lattice. The left column (a,e) shows the interference pattern of the array waves forming the lattice, as they exit the nonlinear crystal. The second column (b,f) shows the power spectra of the input incoherent probe beam (wide homogeneous circle of illumination) illuminating the lattice, with the far-field of the array-forming waves (the sharp peaks) marking the corners of the respective first Brillouin zones (the incoming plane waves define the corresponding Bragg angles). The third row (c,g) shows the text-book calculated picture of the extended Brillouin zones of the respective lattices. And finally, the right column (d,h) shows the experimental far-field picture of the incoherent light emerging from the lattice, depicting the first four Brillouin zones of the lattices, and marking the edges of every zone. These experimental pictures form by Bragg reflections from the regions near the boundaries of the Brillouin zones of each lattice, resulting in dark stripes in the power spectrum of the probe beam at the boundaries. We emphasize that the probe beam intensity in these experiments is much lower than the intensity of the array-forming waves (~1:10), and no external voltage is applied to the crystal while being probed.

Hence, Fig. 2 represents effects caused by linear propagation of the incoherent probe beam in linear photonic lattices.

Next, we examine the spatial transmission spectrum of an incoherent probe beam propagating in a hexagonal (trigonal) lattice with positive and negative defects. The positive defect is created by launching an additional beam (several lattice periods wide), which is co-propagating with the array-forming waves but is mutually incoherent with them. Under a positive bias field, this additional beam induces a positive defect in the lattice, i.e., it increases the index of refraction in the region it illuminates. Likewise, a similar beam induces a negative defect when a negative bias field is applied. The upper row in Fig. 3 (a-c) depicts the near-field photograph of the array-forming waves and the defect-inducing beam as they exit the nonlinear crystal The lower row in Fig. 3 shows the (far-field) power spectrum of the probe beam exiting the hexagonal lattice, without a defect (d), with a positive defect (e), and with a negative defect (f) respectively. For the positive defect, the guided modes (bound states) arise from total-internal-reflection of states occupying the central region of the first Brillouin zone. Consequently, the far-field of these guided modes is clearly apparent as a wide spot in the center of the first Brillouin zone of Fig. 3(e). On the other hand, for a negative defect (for which the average refractive index is lower in the guiding region), waveguiding arises solely from Bragg reflections, with no contribution from total-internal-reflections. As a result, the central region of the Brillouin-zone picture is empty [not populated; the central "hole" in Fig. 3(f)], and the guided modes are modes whose momentum arises from the vicinity of the edge of the first BZ [bright concentric ring in Fig. 3(f))].

Finally, when the photonic lattice is nonlinear, the underlying self-focusing (or self-defocusing) interaction among the Bloch states results in energy exchange between these lattice eigen-modes. This interaction facilitates a method for distinguishing between the regions of normal and anomalous diffraction (dispersion) of the underlying lattice, and mapping out the boundaries between dispersion of opposite signs. In these experiments, the probe beam propagates *nonlinearly* in the photonic lattice, and it drives the nonlinear interaction by inducing a broad defect in the lattice structure, which in turn causes energy exchanges between Bloch states. To do this efficiently, the probe beam intensity must be much higher than in the experiments described by Figs. 2 and 3. In this set of experiments with nonlinear lattices, we use probe beam intensities which are roughly one-half of the intensity of the lattice-forming array waves. Typical experimental results are shown in Fig. 4. The excitation (far-field) power spectrum of the incoherent probe-beam is shown in 4(a). When the crystal is biased with a positive field (self-focusing nonlinearity), the wide probe beam induces a wide (and shallow) positive defect in the lattice (deeper potential / increased refractive index). The presence of such a defect causes Bloch waves from anomalous diffraction regions (negative curvature of the band structure) to transfer power to Bloch waves arising from normal diffraction regions. Indeed, Fig. 4(b) clearly reveals the regions of normal diffraction, which have a considerably higher intensity than the low-intensity regions of anomalous diffraction. In a similar fashion, a negative bias field results in a self-defocusing nonlinearity, through which the incoherent probe beam induces a wide negative defect. This negative defect causes energy transfer from normal-diffraction bloch waves to anomalous-diffraction states, highlighting the higher intensity

(anomalous diffraction) regions in Fig. 4c, adjacent to the lower intensity (normal diffraction) regions in the same far-field power spectrum.

In conclusion, we have shown two different techniques for linear and nonlinear Brillouin zone spectroscopy of photonic lattices, with or without defects. Both methods rely on probing the lattice with an incoherent probe beam and visualizing the far-field power spectrum of the light exiting the photonic structure. The linear method relies on Bragg reflections, resulting in dark stripes along the edges of Brillouin zones, thus marking the edges of every zone. The underlying linear mechanism arises from energy transfer between waves propagating at the close vicinity of the Bragg angles (within the Bragg mismatch), hence the preferential marking of the edges of the various Brillouin zones. Our method of **linear** Brillouin zone spectroscopy is related to the *Kikuchi patterns* (also known as Kossel lines) that have been observed in atomic structures (crystals) through electron microscopy since the 1920's [17,18]. Similar features have been observed in neutron scattering experiments, also in atomic lattices [19]. Finally, recent experiments with stimulated Raman scattering of matter waves in Bose-Einstein condensates (BEC) have also shown features of the extended Brillouin zone map of atomic lattices [20]. However, in the BEC case, the experiments revealed the different populations of the various bands (relying on different band energies), rather than marking the borders between bands through Bragg diffractions (as in our experiments and in those with the Kikuchi patterns). For this reason, the technique used in BEC cannot distinguish between bands that are not separated by a complete gap [20]. Altogether, however, we emphasize, that to the best of our knowledge no such methods for Brillouin zone spectroscopy have ever been demonstrated in photonic structures.

In contrast to our linear Brillouin zone spectroscopy method (which has similarities to in electron microscopy and neutron scattering), our **nonlinear method** adds new features that have never been observed in any system. Namely, our **nonlinear** Brillouin zone spectroscopy method marks the areas of normal and anomalous dispersion (diffraction; effective mass, etc.) wherever they occur within the Brillouin zones of the lattice. This method relies on the nonlinear propagation of the probe beam in the photonic lattice. The probe induces (in real time) a broad defect in the lattice, which results in energy exchange between Bloch waves. During a self-focusing interaction, Bloch waves from anomalous diffraction regions transfer power to Bloch waves arising from normal diffraction regions, and vice-versa for a self-defocusing interaction. The underlying physics behind the nonlinear energy exchange among Bloch states has been intriguing researchers since the famous Fermi-Pasta-Ulam era [11], and its fine details are still not fully understood. Moreover, The ideas underlying our **nonlinear** Brillouin zone spectroscopy method most probably cannot be introduced to electron microscopy or neutron scattering, simply because the latter processes are fundamentally linear. In particular, it would be very difficult to observe nonlinear interactions between electron (or neutron) waves in a crystal and keep the crystalline structure intact. Nevertheless, our nonlinear technique does have immediate relevance to other nonlinear periodic systems (e.g. dynamics in BEC, plasma, sound, etc.), many of which are now being explored in the context of "discrete" solitons [14].

Finally, in addition to characterizing fully periodic photonic lattices, we also used these techniques to characterize photonic lattices with positive and negative defects, observing the far-field power spectrum of their guided and radiation modes. These defect-bearing lattices are akin to photonic crystal fibers and related photonic structures. We have demonstrated our techniques experimentally in optically-induced lattices; nevertheless, the Brillouin zone spectroscopy method is general and can be used for any periodic optical structure.

**Figures**

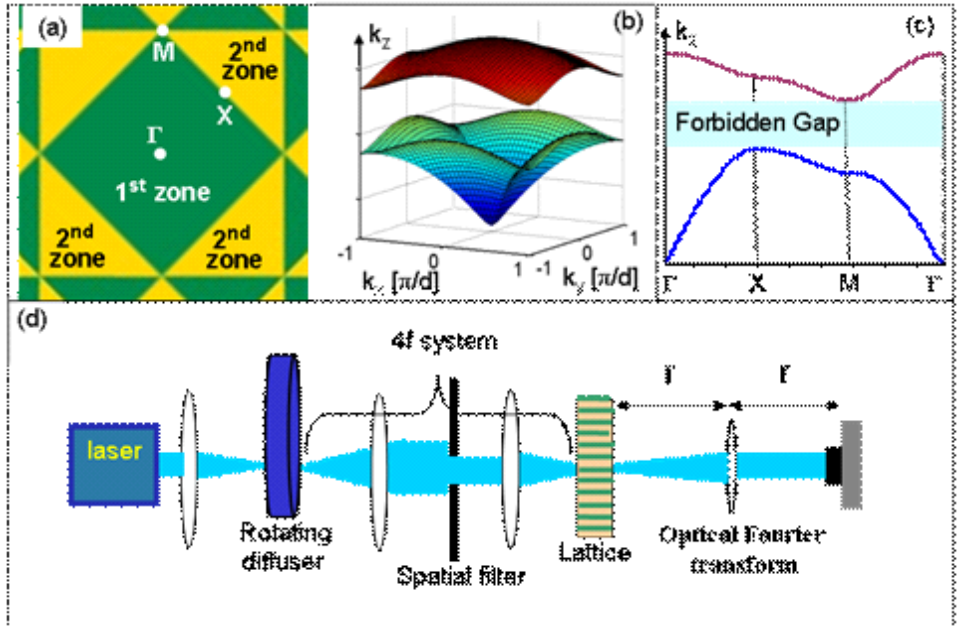

**Figure 1:** **(a) First and second Brillouin zones of a 2D square lattice with the high-symmetry points (Γ, X, and M) marked with white dots. (b)** First two bands of the transmission spectrum of a 2D square lattice. **(c) Dispersion curves between the symmetry points of the first two bands. Negative curvature in these curves corresponds to normal diffraction regions. (d) Diagram of our setup for Fourier-space spectroscopy of photonic lattice.**

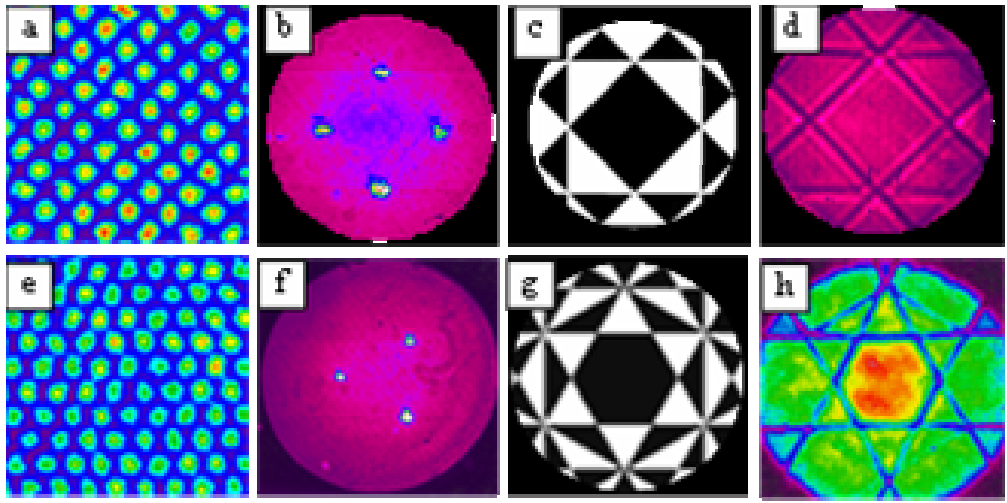

**Figure 2: Experimental linear mapping of the edges of Brillouin zones of square and hexagonal lattices. (a)** Interference pattern of the array waves forming the square lattice at the crystal input face. **(b)** Fourier spectrum of the probe (broad circle) and lattice forming beams (four sharp peaks) at the input face. **(c)** Calculated extended BZ scheme of a square lattice. **(d)** True experimental picture depicting the Fourier spectrum of the probe beam at the output face of the crystal with induced 2D square lattice. **(e)-(h)** Same as a-d with a hexagonal lattice.

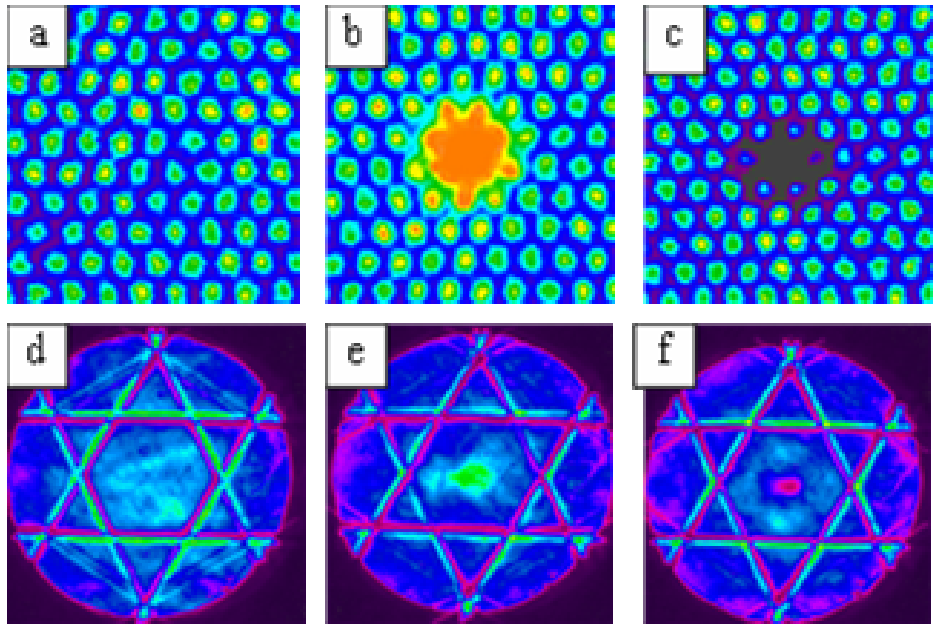

Figure 3: Defect modes in a hexagonal lattice. Interference pattern of the waves forming the hexagonal lattice at the crystal input face (a) with no defect, (b) with a positive defect, and (c) with a negative defect. Experimental Fourier spectrum of the probe beam at the output face of the crystal with induced 2D hexagonal lattice. (d) with no defect, (e) with a positive defect, and (f) with a negative defect.

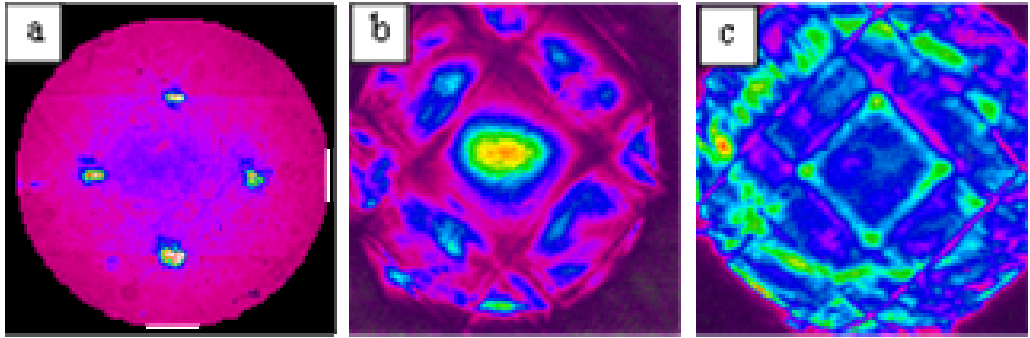

Figure 4: Nonlinear dispersion mapping of a square lat1tice. (a) Fourier spectrum of the probe beam (circle) and of the lattice forming beams (four dots) at the input. Experimental Fourier spectrum of the probe beam at the output face of the crystal with the induced 2D square lattice under (b) self-focusing nonlinearity and (c) self-defocusing nonlinearity